\documentclass{nic-series}

\usepackage[hypertexnames=false]{hyperref}

\DeclareFieldFormat*{volume}{\mkbibbold{#1}} 
\DeclareFieldFormat*{title}{\mkbibemph{#1}} 
\DeclareFieldFormat*{journaltitle}{#1}

\newcommand{\LA}{\left \langle}
\newcommand{\RA}{\right \rangle}

\newcommand{\barechiralcondinlattice}{\left \langle \Bar{\psi}\,\psi \right \rangle}
\newcommand{\barechiralcond}{M_\ell}
\newcommand{\barechiralsusc}{\chi_\ell}

\newcommand{\Ob}{\mathcal{O}}
\newcommand{\partitionfunction}{\mathcal{Z}}
\begin{document} 

\newrefsection[article-example]

\title{A parameter independent analysis of the QCD chiral phase transition and its universal critical behaviour}

\author{Jishnu Goswami  \and 
        Frithjof Karsch  \and \\
        Sabarnya Mitra  \and 
        Christian Schmidt 
        }

\authortoc{J. Goswami, F. Karsch, S. Mitra, C. Schmidt}

\institute{Fakult\"at f\"ur Physik, Universit\"at Bielefeld, D-33615 Bielefeld, Germany\\
         \email{\{jishnu, karsch, smitra, schmidt\}@physik.uni-bielefeld.de}
          }

\maketitle

\begin{abstracts}
 
We make use of unique properties of scaling functions to estimate
the chiral phase transition temperature and corresponding universal critical parameters of this phase transition directly from (2+1)-flavor QCD simulations with highly improved staggered fermions on lattices with temporal extent $N_\tau=8$.
Working with an improved chiral order parameter for quantifying chiral symmetry breaking, we analyze the finite-volume dependence of this observable for a wide range of lattice volumes and light quark masses, and quantitatively estimate the deviations from the expected universal scaling behaviour as a function of the light-to-strange quark mass ratio.
\end{abstracts}

\section{Introduction}
The chiral phase transition in Quantum Chromodynamics (QCD) across the high temperature plasma and low temperature hadronic phases in the limit of vanishing up and down quark masses, is instrumental for constraining the QCD critical point in the QCD phase diagram \cite{Ding:2024sux,Goswami:2025wtj,Schmidt:2025ppy} as well as, for providing useful insights on the fate of the $U(1)$ axial anomaly at this phase transition temperature.
Even though existing lattice studies \cite{HotQCD:2012vvd,Ding:2020xlj} reveal the restoration of the spontaneously broken $SU(2)_L \times SU(2)_R$ chiral symmetry along with explicit $U(1)_A$ breaking at this temperature, the influence of this anomaly on the universal behaviour of the chiral phase transition, in particular the nature of its breaking and its extent of effect on the physical QCD observables continues to remain an open avenue to be explored. It is therefore crucial to provide further evidences in support of the universal critical nature of this phase transition which correspond to the three-dimensional $O(4)$ class in the chiral limit resembling vanishing light quark masses \cite{Pisarski:1983ms}. 
While several studies are already existent in literature exploring this chiral universal behaviour, a majority of these numerical studies start with the presumption that the transition belongs to the three-dimensional $O(4)$ universality class at the very onset. Hence for subsequent benchmarking of lattice results, these studies construct the corresponding universal scaling ansatz using the $O(4)$ critical exponents. This makes a direct lattice QCD determination of the critical parameters influencing the QCD chiral phase transition obviously more desirable in this paradigm. 

Having this as the main aim and purpose of this work here, we systematically approach towards a direct lattice QCD based calculation of the critical universal parameters related to the QCD chiral phase transition. 
Although the present study is limited to a fixed lattice spacing,
we rather focus on a detailed analysis of finite volume and non-vanishing quark mass effects on chiral observables that eventually will allow a direct determination of universal amplitudes and critical exponents in QCD, without relying on any form of a priori assumptions related to the underlying universality class. 
Working within the domain of zero density here, we briefly introduce in this proceedings, an improved order parameter and discuss its universal features prior to elaborating its finite-size scaling (FSS) dependence within the framework of (2+1)-flavor, lattice regularized QCD based on lattices with temporal extent $N_\tau=8$.
We reported on this ongoing research project previously in Refs.~\cite{Mitra:2024mke,Mitra:2025aeu,Mitra:2025hsk}.

\section{Theory}
\label{attig_sec_motive}

\subsection{Improved order parameter and its susceptibility}

We outline here, some theoretical background for our study of the universal properties of thermodynamic observables close to the chiral phase transition in (2+1)-flavor QCD. The relevant quantities of interest are the $2$-flavor light quark chiral condensate, $\barechiralcond$, and the corresponding chiral susceptibility, $\barechiralsusc$, 
  \begin{equation}
      \barechiralcond = \frac{m_s}{f_K^4} \barechiralcondinlattice_\ell, \hspace{1cm}
      \barechiralsusc = m_s \,\frac{\partial \barechiralcond}{\partial m_\ell} \; .
      \label{eq:basics}
  \end{equation}
Here, we introduced factors of the strange quark mass, $m_s$, as multiplicative renormalization factors and the kaon decay constant, $f_K$, is used to obtain dimensionless quantities, which can be directly obtained from lattice calculations. The former is kept at its physical value, by tuning the mass of the pseudoscalar meson, $\eta_{\bar{s}s}$\footnote{The physical $\eta_{\bar{s}s}$ is a mixture of flavor-octet and flavor-singlet states.}, to its physical value on the line of constant physics (LCP) as detailed in \cite{HotQCD:2014kol}.
On a $4$-d space-time lattice with spatial and temporal extents $N_\sigma$ and $N_\tau$, and lattice spacing $a$, chiral observables are derived using $m_\ell$-derivatives of the free energy density, $f=-T\ln \partitionfunction/V$. Here,
\begin{eqnarray}
	\partitionfunction(N_\sigma,N_\tau,m_\ell,m_s) = \int DU \; e^{-S_g[U]}\;\big[\det D_\ell(U,m_\ell)\big]^{1/2}
	\times\big[\det D_s(U,m_s)\big]^{1/4}   ,
	\label{eq:partitionfunction}
\end{eqnarray}
is the QCD partition function with $D_\ell$ and $D_s$ denoting the light and strange quark staggered Dirac operators respectively on a system of volume $V$, with gauge action $S_g$ of gauge fields $U$. The volume $V$ and temperature $T$ are defined as,
\begin{equation*}
    V=(aN_\sigma)^3 \quad,\quad T = (aN_\tau)^{-1}\;,
\end{equation*}
With this we obtain
the light quark chiral condensate in $2$-flavor normalisation,
\begin{equation}
\LA \bar{\psi}\psi \RA_\ell
= -
\frac{\partial f}{\partial m_\ell} = 
\frac{1}{2} \frac{1}{N_{\sigma}^3 N_{\tau}} 
\Big \langle \,{\rm Tr} \, D_\ell^{-1} \Big \rangle\,.   
\label{eq:barechiralcond}
\end{equation}
and the subsequent chiral susceptibility, $\chi_\ell=\chi_{\ell,disc}+\chi_{\ell,conn}$, which comprise disconnected and connected parts, $\chi_{\ell,disc}$ and $\chi_{\ell,conn}$,
\begin{align}
    \chi_{\ell,disc} &= \frac{1}{2} \frac{1}{N_{\sigma}^3 N_{\tau}} \left[\LA \left({\rm Tr} \, D_\ell^{-1}\right)^2 \RA - \Big \langle {\rm Tr} \, D_\ell^{-1} \Big \rangle^2\right] 
    \label{eq:disconn susc}\\ 
    \chi_{\ell,conn} &= -\frac{1}{2} \frac{1}{N_{\sigma}^3 N_{\tau}} \Big \langle {\rm Tr} \, \left(D_\ell^{-1}\,D_\ell^{-1}\right) \Big \rangle\;.\label{eq:conn susc}
\end{align}
The multiplicatively renormalised chiral condensate $M_\ell$ introduced in Eq.\eqref{eq:basics} still contains $\Ob(a^{-2})$ ultraviolet divergent terms, which inhibits a true continuum limit behaviour of this observable. In order to therefore have a well-defined value of this order parameter in the continuum limit, one introduces additive corrections to this chiral condensate. This is,
for instance, achieved by subtracting a suitable
fraction of $\chi_\ell$ from $M_\ell$ \cite{Unger:2010wcq},
 \begin{equation}
     M(T,H,L) = \barechiralcond (T,H,L) - H\barechiralsusc(T,H,L)\; \;.
     \label{eq:improved-M}
 \end{equation}
Here, $H\equiv m_\ell/m_s$ denotes the ratio of light-to-strange quark masses and $L=N_\sigma/N_\tau$ denotes the system size in units of the inverse temperature.

\subsection{Scaling functions and critical behaviour}

Close to the chiral critical point, 
the free energy density and the chiral  observables deduced from it can be written as sum of singular and sub-leading parts where the latter comprises corrections-to-scaling and genuine regular terms, 
which are analytic and can be expressed in terms of a Taylor series in the vicinity of the critical point. The singular terms are non-analytic functions in the vicinity of the critical point and can be expressed in terms of scaling functions,
  \begin{eqnarray}
      &&M_\ell(T,H,L)  = h^{1/\delta} \, f_G(z,z_L) 
      + M_{\ell,\,{\rm sub-lead}}(T,H,L)\; , 
      \label{eq:barechiralcondregscaling}  
      \\
      &&\chi_{\ell}(T,H,L)  = \frac{1}{\delta}\,h^{1/\delta-1} f_{\chi}(z,z_L) +\chi_{\ell,\,{\rm sub-lead}}(T,H,L)\; ,
      \label{eq:barechiralsuscregscaling} 
      \\
      &&M(T,H,L)  = h^{1/\delta} \, f_{G\chi}(z,z_L) 
      + M_{{\rm sub-lead}}(T,H,L)\; .
      \label{eq:improvedchiralcondregscaling}       
      \label{eq:scaling of observables}
  \end{eqnarray}
Here, the scaling functions $f_G, f_\chi$ and $f_{G\chi}=f_G-f_\chi$ appear having as arguments, the scaling variables $z=t\,h^{-1/\Delta},z_L=l\,h^{-\nu/\Delta}$, where $\Delta=\beta\delta$. While the parameters $\beta,\delta$ and $\nu=\beta\left(1+\delta\right)/3$ are the critical exponents of the relevant three-dimensional universality class, the quantities $t,h,l$ are defined as 
\begin{equation}
    t=\frac{\tau}{t_0}\quad, \quad h = \frac{H}{h_0} \quad , \quad l = \frac{l_0}{L}
\end{equation}
with reduced temperature $\tau=(T-T_c)/T_c$ having the chiral phase transition temperature $T_c$ and non-universal system-dependent constants $t_0,h_0,l_0$. 
In the following as well as in the figures, we will use the notations, $z_0=h_0^{1/\Delta}/t_0$ and $z_{L,0}=l_0\,h_0^{\nu/\Delta}$. \footnote{One can also write $z=z_0\,z_b\;,\;z_L=z_{L,0}\,z_{L,b}$, \quad where, \quad $z_b = \tau\,H^{-1/\Delta}\;,\;z_{L,b}=H^{-\nu/\Delta}/L$\;.}

Alongside having a well-defined continuum limit, it is evident from Eq.~\eqref{eq:improved-M} that there is no linear light quark mass dependence in the improved order parameter $M$. Instead the leading order dependence is cubic, $\Ob(H^3)$, which predominantly constitutes the analytic regular correction to the singular part of this observable in the chiral limit, $H \to 0$. 

The direct relation of the 
improved order parameter $M$ to 
the scaling function $f_G$ and $f_\chi$
makes $M$ a suitable observable for studying the critical regime and singular behaviour, although the subtraction of a susceptibility potentially introduces larger finite-volume effects. 
\begin{figure}[t]
\includegraphics[height=0.37\textwidth]{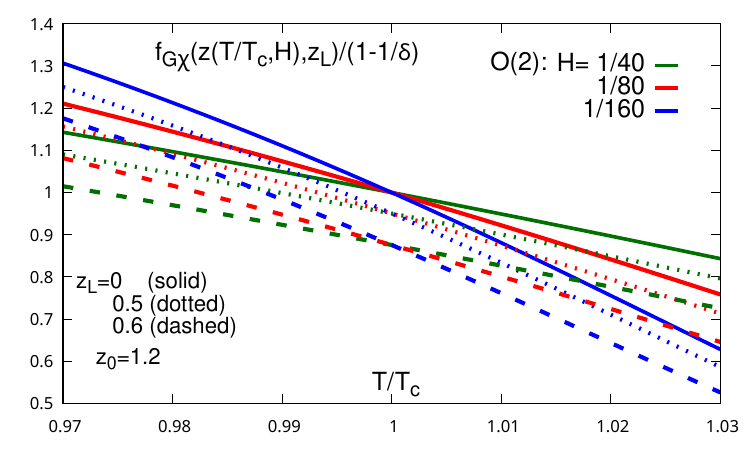}
\includegraphics[height=0.37\textwidth]{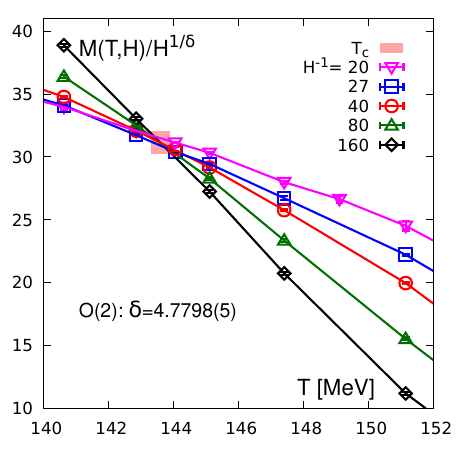}
\caption{
(Left) The $O(2)$ scaling function $f_{G\chi}(z)$ versus $T/T_c$ for an arbitrary, $z_0=1.2$ for infinite ($z_L=0$) and finite-volume ($z_L \neq 0$) limits, shown by solid and dotted set of curves for different $H$ values. (Right) Lattice Data of $M_{resc}=H^{-1/\delta}\,M$ as function of $T$ using $\delta=4.7798$ of $O(2)$ universality class. }
\label{fig:fGfGchi}
\end{figure} 
The universal FSS functions $f_{G}(z,z_L)$ and $f_{G\chi}(z,z_L)$
have been determined for several universality classes including $O(2)$, which is shown in Fig.~\ref{fig:fGfGchi}. These scaling functions are related,
\begin{equation*}
    f_{G\chi}(z,z_L) = \left(1-\frac{1}{\delta}\right)\,f_G(z,z_L) 
    + 
    \frac{z}{\Delta}\,\frac{\partial f_G(z,z_L)}{\partial z} 
    +
    \frac{\nu z_L}{\Delta}\,\frac{\partial f_G(z,z_L)}{\partial z_L}\;.
\end{equation*}
For a recent parametrization, see for instance Ref.~\cite{Karsch:2023pga}. It is evident that at the chiral critical point
$(z,z_L)=(0,0)$, the scaling function is
independent of $H$,
$f_{G\chi}(0,0)=\left(1-1/\delta\right)$, thus defining a unique intersection point for various fixed-$H$ curves, as seen in Fig.~\ref{fig:fGfGchi}. 
We obtain similar results from lattice simulations for the rescaled improved order parameter $M_{resc}$ \cite{Mitra:2024mke}, 
\begin{equation*}
    M_{resc}=H^{-1/\delta}\,M = h_0^{-1/\delta}\,f_{G\chi}(z,z_L)+M_{resc,{\rm  sub-lead}}
\end{equation*}
 shown in the right plot of Fig.~\ref{fig:fGfGchi}. On using the value of $\delta=4.7798$ of $O(2)$ universality class, we find the smaller-than-physical quark mass curves, i.e. $H \leq 1/40$, intersect at a unique point in $T$ within the working level of precision, which agrees well with previous similar determinations, $T_c(N_\tau=8)=143.7$  MeV \cite{Ding:2024sux}. 
 This agreement becomes enhanced in the chiral limit, or close to vanishing light quark masses. 
 
As can be seen in Fig.~\ref{fig:fGfGchi}-(left),
the location of the intersection point remains at $T=T_c$ also for non-vanishing $z_L$. However, unlike the infinite-volume limit corresponding to $z_L=0$, the value $f_{G\chi}(0,z_L)$ ceases to provide a straightforward estimate of $\delta$, because of the not well known parametric form of the $z_L$-corrections it receives at finite volumes. 
For this reason, we started to perform a more systematic analysis of the finite volume dependence of chiral observables.
We will present some results from this ongoing study in the next section.

\section{Finite-volume dependence and effects in the chiral order parameter}

In order to arrive at a direct determination of universal critical exponents from lattice QCD calculations
performed at non-vanishing values of the 
quark masses, the need for a better understanding of universal and non-universal finite-volume
effects contributing to chiral observables is very important.

\begin{figure}[t]
\includegraphics[scale=0.49]{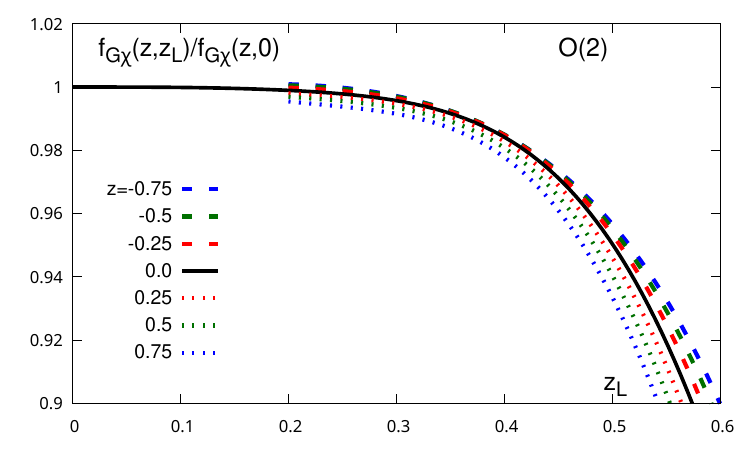}
\includegraphics[scale=0.5]{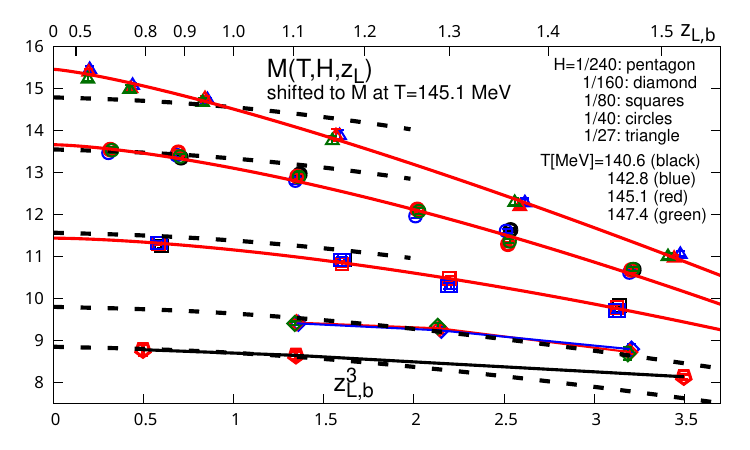}
\caption{(Left) The ratio of finite and infinite volume scaling function $f_{G\chi}$ as function of the FSS variable $z_L$ for different values of $z$ including $T=T_c(z=0)$ (black solid curve) for the three-dimensional $O(2)$ universality class. (Right) The lattice results for improved 
order parameter, $M$ vs. bare size scaling variable $z_{L,b}^3 \sim 1/V$ for quark masses, $1/240 \leq H \leq 1/27$ at four values of $T \sim 141, 143, 145$ and $147$ MeV. }
\label{fig:fGchiratfiniteV-and-temp-dep}
\end{figure}
Universal scaling functions for the order parameter and the order 
parameter susceptibility have been analyzed in quite some
detail for the three-dimensional $O(2)$ universality class (see e.g. \cite{Engels:2001bq}), although an explicit 
parametrization for FSS functions in this universality class has been given in terms of a Taylor series in $z$ and $z_L$ \cite{Karsch:2023pga},
\begin{eqnarray}
f_{G}(z,z_L) &=& f_{G}(z) + 
\sum_{n=0}^{5}\sum_{m=3}^{8} a_{nm} \,z^n\, z^m_L
\; , 
\label{eq:fgchiV}
\end{eqnarray}
in the range $-1$ $\leq$ $z$ $\leq$ $2$ and $0.4$ $\leq$ $z_L$ $\leq$ $1.0$, 
with coefficients $a_{nm}$ \cite{Karsch:2023pga}.  
In the infinite volume limit, $z_L=0$, the volume-dependent term of the 
scaling functions obviously vanishes and we obtain the standard one-parameter form of the infinite volume scaling function $ f_G(z)$. While probing the $O(2)$ FSS function $f_{G\chi}$ as function of $z_L$ for temperatures close to $T_c$ corresponding to $|z| < 1$ as shown in Fig.~\ref{fig:fGchiratfiniteV-and-temp-dep}, we observe a small finite-volume effect 
also 
for $z_L \leq 0.4$.
This corresponds to $z_{L,b}=H^{-\nu/\Delta}/L \leq 1$ at $T_c$, considering the value of $z_{L,0}=0.39$ \cite{Karsch:2023pga}. 
These finite-volume effects
are small
also for other $T$ values close to $T_c$. 
For $z_{L,b} > 1$ finite-volume effects rapidly become larger and one encounters
greater effects, for instance around $10\%$ for $z_{L,b} \sim 1.5 \equiv z_L=0.6$. 
Also seen in Fig.~\ref{fig:fGchiratfiniteV-and-temp-dep} is that close to $T_c$ the
approach to the infinite volume limit is faster
than $1/V\sim z_L^3$. 
Quantifying the parametric dependence of this $z_L$-dependence is part of
our ongoing FSS study.
The relevant lattice results in the range of $z_{L,b}^3\le 3.5$ are shown in Fig.~\ref{fig:fGchiratfiniteV-and-temp-dep}-(right).  
These results are, to a good extent, consistent with
the $O(2)$ spin model scaling function predictions close to the infinite-volume limit. 
Deviations from this scaling behaviour are observed
for smaller volumes resulting in enhanced finite-volume effects as well as for larger quark masses, i.e. away from the chiral limit for all these four temperatures studied.
\begin{figure}[htbp]
\includegraphics[scale=0.495]{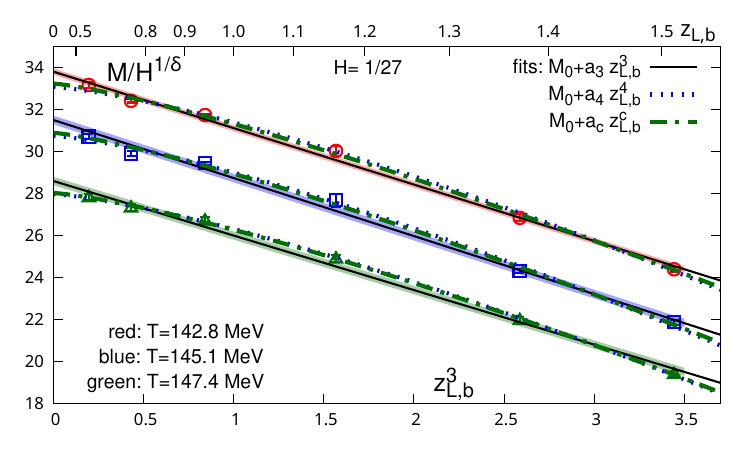}
\includegraphics[scale=0.495]{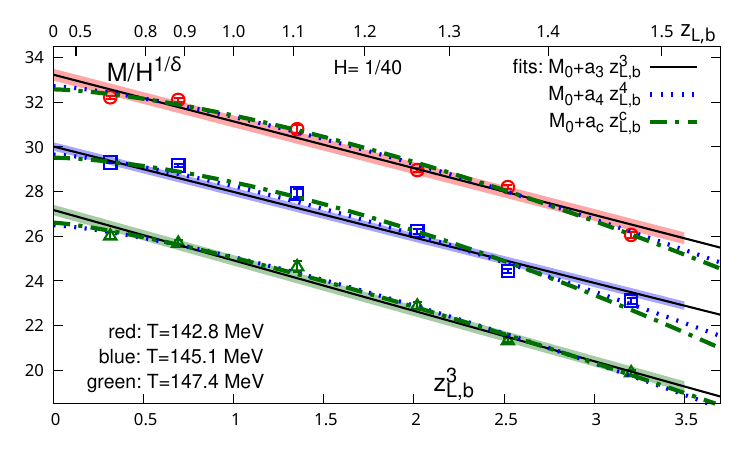}
\includegraphics[scale=0.495]{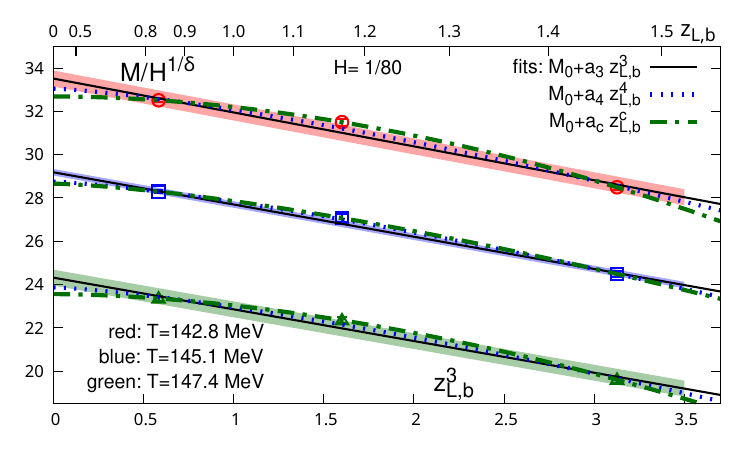}
\includegraphics[scale=0.495]{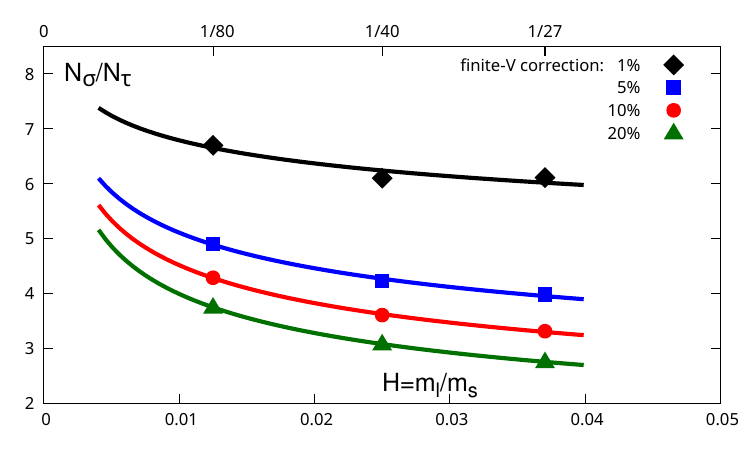}
\caption{The scaled order parameter, $M/H^{1/\delta}$, for 
three values of $H=1/27, 1/40$ and $1/80$, versus inverse volume $\sim z_{L,b}^3$. (Bottom right) Different finite-volume curves with various magnitudes of finite-size effects where the aspect ratio, $N_\sigma / N_\tau$, of $N_\sigma^3 \times N_\tau$ lattices is plotted as a function of quark masses, $H=m_\ell/m_s$   
}
\label{fig:Mscaled}
\end{figure}

\begin{table}[ht]
\centerline{
\begin{tabular}{|c|c|c|c|c|}
\hline
$H=m_\ell/m_s$ & $N_\sigma$ & $z_{L,b}=H^{-\nu/\Delta}/L$ & $z_{L,b}^3$ & $L=N_\sigma/N_\tau$\\[.1ex]
\hline\hline
1/27 & 52 & 0.5808 & 0.1960 & 6.50 \\[0.45ex]
1/40 & 52 & 0.6805 & 0.3151 & 6.50 \\[0.45ex]
1/80 & 56 & 0.8356 & 0.5834 & 7.00 \\[0.45ex]
1/160 & 80 & 0.7734 & 0.4626 & 10.00\\[0.45ex]
1/240 & 92 & 0.7919 & 0.4966 & 11.25\\[0.45ex]
\hline
\end{tabular}
}  
\caption{\label{tab:setup}
  Table showing the present working setup of our analysis on $N_\sigma^3 \times 8$ lattices involving quark masses, $H$ and the respective highest lattice volumes $N_\sigma$ used with $L=N_\sigma/8$, providing corresponding values of $z_{L,b}$.}
\end{table}

For $H=1/40,1/80$ including the physical quark mass, $H=1/27$, we outline the volume dependence of the rescaled order parameter, $M_{resc}=M/H^{1/\delta}$ for $T=143,145$ and $147$ MeV in Fig.~\ref{fig:Mscaled}. It has been argued \cite{Mitra:2025hsk} 
that the leading order dependence is closer to fourth order in $z_{L,b}$ i.e. $z_{L,b}^4 \sim V^{-4/3}$, rather than a simple linear dependence on inverse volume $V^{-1} \sim z_{L,b}^3$. Detailed discussion about these fits are already mentioned in \cite{Mitra:2025hsk} and are also part of our ongoing work in progress. For studying the finite-volume effects for our working quark masses on different lattices, we also plot trajectories of constant finite-volume effects on $L$-$H$ plane in Fig.~\ref{fig:Mscaled}-(bottom right).
This clearly suggests that in order to mitigate finite-size effects, one  essentially needs to consider larger lattice volumes for studying the critical behaviour which  
requires approaching the chiral limit, by considering smaller values of $m_\ell$ or $H$ (presently, $H=1/240$ is the smallest quark mass considered in our ongoing work).
Parameters used in our current work in progress
on $N_\tau=8$ lattices are reported in Table \ref{tab:setup}. 
\begin{figure}[t]
\begin{center}
\includegraphics[width=0.7\textwidth]{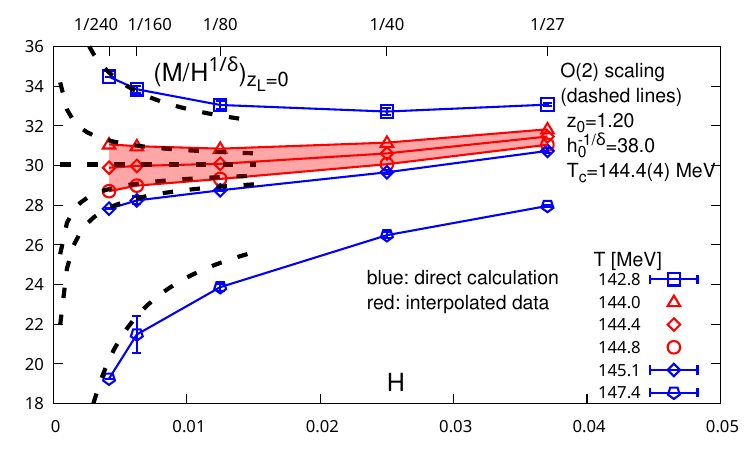}
\caption{The infinite volume extrapolated, scaled order parameter versus 
$H$ for three values of the temperature 
(blue) and interpolated values (red) used
to estimate the chiral phase transition temperature on lattices with temporal extent $N_\tau=8$. 
Dashed lines show the infinite volume, asymptotic scaling results using the 
non-universal parameters given in the figure.
}
\label{fig:Tcestimate}
\end{center}
\end{figure}

Lastly, we outline the present updated form of the infinite-volume extrapolated result of the rescaled order parameter, $M_{resc}$ as function of quark mass, $H$ in Fig.~\ref{fig:Tcestimate}. Taking into account, the finite-volume analysis of $H=1/160$ and new $H=1/240$ results performed on $80^3 \times 8$ and $92^3 \times 8$ lattices (see Table \ref{tab:setup}), we find that the finite-volume corrected estimate of $T_c$ increases from previous $143.7$ MeV to $144.4(4)$ MeV. This is understood  from the asymptotic behaviour of the scaling function, 
$f_{G\chi}$ in the limit $z\rightarrow \pm \infty$,
which becomes relevant when
approaching
the chiral limit for $T\ne T_c$.  
In the chiral limit the improved order parameter normalized by $h^{1/\delta}$ diverges for $T<T_c$, vanishes smoothly for $T>T_c$ and converges to a non-vanishing finite value at $T=T_c$. 
While direct lattice data points are present only for $T=142.8,145.1$ and $147.4$ MeV, we interpolate this behaviour in between $142.8$ and $145.1$ MeV for three more temperatures as outlined here by observing the qualitative change of $M_{resc}$ for these two temperatures. Importantly, this constrains the new estimate of $T_c$ by implying $142.8 \; {\rm MeV} < T_c < 145.1 \; {\rm MeV}$. This is still a work in progress, with lattice simulations ongoing at $H=1/160$ and the new data point $H=1/240$.

\section{Conclusions}
 
We have shown that, the rescaled order parameter evaluated as function of $T$
for different values of the light-to-strange quark mass ratio $H=m_\ell/m_s$ exhibits the feature
of having a unique intersection point in the chiral limit. More precisely, this is the limit where the light quark mass $m_\ell$ or $H$ is small enough so that contributions from sub-leading terms can be neglected. 
However, in order to arrive at the point, where these calculations reach a sufficient accuracy to allow for quantitative determination of $\delta$
thereby enabling a clear distinction between universality classes relevant 
for the analysis of the chiral phase transition, one needs to collect more 
information on the $H$ and $L$ dependence of 
the rescaled order parameter close to the chiral limit. A more detailed analysis of finite volume as well as cut-off effects will be needed. The latter were not subject of the discussion presented here. In fact, this is work in progress.

We presented first results from an ongoing study
of finite-size effects in calculations of an improved
order parameter for chiral symmetry breaking in $(2+1)$-flavor QCD, in which we attempt to quantify the finite-size effects as function of lattice volume and the light quark masses for a given temperature close to the chiral phase transition. Our calculations on lattices
with fixed lattice cut-off corresponding to a temporal extent of $N_\tau=8$ show that precision studies of the order parameter require aspect ratios for the lattice size in units of temperature, $N_\sigma/N_\tau\ge 6$ at physical values of the 
quark masses, which increases to about $11$ for the currently used
smallest quark mass ratio $H=m_\ell/m_s=1/240$. This is currently a work in progress.
We also show that the improved order parameter $M$
receives corrections to the leading order universal
critical behaviour from corrections-to-scaling and/or
regular terms for $m_\ell/m_s>1/160$. These corrections stay below the 10\% level at the physical value of the light-to-strange quark mass ratio $H=1/27$.
A more detailed, quantitative analysis of corrections
to leading order universal scaling does require
further calculations at small values of  $m_\ell/m_s$\,,
which are ongoing.

\section*{Acknowledgements}
This work is supported by the Deutsche Forschungsgemeinschaft, German Research Foundation, Proj. No. 315477589-TRR 211. We sincerely acknowledge the computing time made
available on clusters Noctua2 and Otus at the NHR Center Paderborn Center for
Parallel Computing (PC2),  supercomputer JUPITER at J\"ulich Supercomputing Centre (JSC) by
the Gauss Centre for Supercomputing e.V.,  EuroHPC supercomputer LEONARDO hosted by
CINECA (Italy) through the EuroHPC Joint Undertaking and Bielefeld HPC-GPU clusters in
Bielefeld University. We also sincerely thank Olaf Kaczmarek, Mugdha Sarkar
and Sipaz Sharma for useful discussions and their contributions to this ongoing research project.

\printbibliography

\end{document}